\documentclass{aipproc}
\usepackage{amsmath,amssymb}
\usepackage{amsfonts}
\layoutstyle{8x11double}

\begin{document}
\title         {Thermal Right-Handed Sneutrino Dark Matter}

\author{F. Deppisch}{
address={School of Physics and Astronomy, University of Manchester, Manchester M13 9PL, UK}
}
\author{A. Pilaftsis}{
address={School of Physics and Astronomy, University of Manchester, Manchester M13 9PL, UK}
}

\begin{abstract}
We discuss the relic abundance of the right-handed sneutrinos in the supersymmetric $F_D$-term  model of hybrid inflation. As well as providing a natural solution to the $\mu$- and gravitino overabundance problems, the model offers the lightest right-handed sneutrino as a candidate for thermal dark matter. The $F_D$-term model predicts a new quartic coupling of purely right-handed sneutrinos to the Higgs doublets that thermalizes the sneutrinos and makes them annihilate sufficiently fast to a level compatible with the current cosmic microwave background data. We discuss this scenario and identify  favourable regions of the parameter space within mSUGRA. 

\end{abstract}

\maketitle

\section{$F_D$-Term Model of Hybrid Inflation}
Hybrid inflation~\cite{Linde}, along with its supersymmetric realizations~\cite{CLLSW}, remains one of the  most predictive models of inflation. In such a scenario, inflation terminates at some critical point when a so called waterfall field acquires a vacuum expectation value~(VEV), thereby ending inflation. Recently, a new supersymmetric hybrid inflationary model was proposed~\cite{GP}, which realizes $F$-term hybrid inflation and includes a subdominant non-anomalous  Fayet-Iliopoulos  $D$-term  that  arises  from  a  U(1) gauge symmetry of the inflation waterfall  sector. It has therefore been called the $F_D$-term  model of  hybrid inflation.  The $F_D$-term model can naturally  accommodate the currently favoured red-tilted spectrum, along  with the actual  value of  the power  spectrum of  curvature perturbations, and the required  number  of $e$-folds.

The $F_D$-term model is defined through the superpotential
\begin{eqnarray}
\label{Wmodel}
	{\cal W}\ &=&\  \kappa\,\widehat{S}\,
        \Big(\widehat{X}_1\widehat{X}_2\:-\:M^2\Big)\ 
	+\ \lambda\,\widehat{S} \widehat{H}_u  \widehat{H}_d\ 
	\nonumber\\
	&+&\ \frac{\rho_{ij}}{2}\,\widehat{S}\, \widehat{N}_i\widehat{N}_j\ 
	+\ h^{\nu}_{ij} \widehat{L}_i \widehat{H}_u\widehat{N}_j
 	+\ W_{\rm MSSM}^{(\mu = 0)}\; ,
\end{eqnarray}
where  $\widehat{S}$  is  the  gauge-singlet inflaton  superfield  and $\widehat{X}_{1,2}$  is  a  chiral  multiplet pair  of  the  so-called waterfall fields which have  opposite charges under the U(1)$_X$ gauge group,  i.e.~$Q (\widehat{X}_1)=-Q  (\widehat{X}_2)=1$.   In addition, $W_{\rm MSSM}^{(\mu  = 0)}$ indicates the  MSSM superpotential without the $\mu$-term. The model Lagrangian is supplemented by soft SUSY breaking terms, which contribute to the scalar potential, leading to an inflaton VEV naturally of \({\cal O}\)~(TeV). 

Constraints on the model can be derived from cosmological inflation  through the number of $e$-folds and the power spectrum and spectral index of curvature perturbations~\cite{GP}. This implies the  upper limit \(\lambda(M_{\rm SUSY}) \lesssim 1.14 (1.82)\times 10^{-2}\) for an inflaton sector with a minimal (next-to-minimal) K\"ahler potential~\cite{Deppisch:2008bp}.

The presence  of the Fayet-Iliopoulos term  in the $F_D$-term model is necessary to approximately break a discrete symmetry in the waterfall sector. The late decays of the GUT-scale  waterfall particles  produce an  enormous entropy  that can reduce  the  gravitino abundance below  the limits imposed by big bang nucleosynthesis (BBN), providing a viable solution to the gravitino overabundance problem~\cite{GP}.

In the $F_D$-term model the $\mu$-parameter of the MSSM can be  generated effectively  when  the scalar inflaton receives a non-zero VEV, $\mu=\lambda\langle S \rangle$. Moreover, the inflaton VEV will also produce an effective Majorana mass matrix as well, $M_N=\rho \langle S \rangle$. As a consequence, the  resulting heavy Majorana neutrinos are expected  to have masses  of order $\mu$. A possible explanation of the observed  baryon asymmetry in the Universe may be obtained by thermal electroweak-scale resonant  leptogenesis~\cite{PU2}.

In this report~\cite{talk}, which summarizes the results of Ref.~\cite{Deppisch:2008bp}, we focus on the properties of the lightest right-handed sneutrino (LRHS) and identify  favourable regions of the parameter space within mSUGRA, for which it becomes a viable candidate for thermal dark matter.

\section{Sneutrino Mass Spectrum}
Ignoring the terms proportional to the small neutrino-Yukawa couplings, the \(6\times 6\) right-handed sneutrino mass matrix ${\cal M}^2_{\widetilde  N}$ is  given in  the weak  basis $(\widetilde{N}_{1,2,3}, \widetilde{N}^*_{1,2,3})$ by
\begin{equation}
	\label{eq:SneutrinoMassMatrix}
	{\cal M}^2_{\widetilde N} = 
	\frac{1}{2} 
	\left(\begin{array}{cc}
		\rho^2 v^2_S      + M^2_{\widetilde  N} & 
		\rho A_\rho v_S   + \rho\lambda v_u v_d \\
		\rho A^*_\rho v_S + \rho\lambda v_u v_d &
		\rho^2 v^2_S      + M^2_{\widetilde  N}
	\end{array} \right),
\end{equation}
where  $v_S = \langle S \rangle$, $v_{u,d} = \langle H_{u,d} \rangle$. Neglecting the possible flavor structure   contained in the soft SUSY-breaking sneutrino mass matrix \(M^2_{\widetilde N}\) and trilinear coupling matrix \(A_\rho\), the  sneutrino spectrum will consist of 3 light (3 heavy) right-handed sneutrinos with masses
\begin{equation}\label{eq:SneutrinoMasses}
	m^2_{\tilde N_{L(H)}} = 
	     \rho^2 v^2_S 
	+    M^2_{\widetilde N}  
	-(+) \left|\rho A_\rho v_S + \rho\lambda v_u v_d\right|.
\end{equation}
All mass terms in (\ref{eq:SneutrinoMasses}) are ${\cal O}(100$--1000)~GeV, so  a  proper  choice  of  model parameters can accommodate a lightest right-handed sneutrino to act as LSP.  Unless  the trilinear coupling \(A_\rho\) is small compared  to \(\mu\), the off-diagonal elements in (\ref{eq:SneutrinoMassMatrix}) will  induce a sizeable  mixing between the  heavy and  light right-handed  sneutrino states,  suppressing the light masses to  values smaller than \((\mu^2+M_{\tilde N}^2)^{1/2}\).
\begin{figure}
\begin{minipage}{\columnwidth}
\centering
\includegraphics[clip,width=1\textwidth]{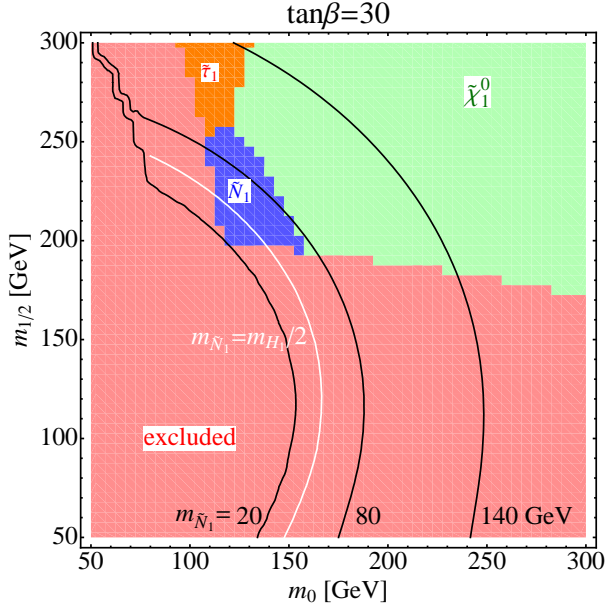}
\end{minipage}
\caption{Allowed  $(m_0,\,  m_{1/2})$  parameter  space for  \(A_0=300\)~GeV, \(\mu>0\), \(\tan\beta=30\)  and \(\lambda=\rho=10^{-2}\).  The black contours show the predicted LRHS mass, while a sneutrino \(\tilde N_1\)/neutralino \(\tilde\chi^0_1\)/stau \(\tilde\tau_1\) LSP is present in the blue/green/orange area. The  red area is  excluded by  direct SUSY  mass searches.   The white contour  is  defined by  the  condition \(m_{\tilde  N_1}=m_{H_1}/2\).}
\label{fig:scans}
\end{figure}

In  Figure~\ref{fig:scans} we plot the LHRS  mass \(m_{\tilde  N_1}\)   as  contours  in  the   mSUGRA  parameter  plane (\(m_0,m_{1/2}\)), for \(\tan\beta=30\), \(A_0=300\)~GeV  and  \(\mu>0\). For the inflaton couplings \(\lambda,\rho\) we simply choose \(\lambda=\rho=10^{-2}\), in accordance with the bounds derived from inflation. The connection between the LRHS mass \(\tilde N_1\) and \(\mu\) generally points towards a low-energy SUSY  spectrum. This coincidentally includes the \(H_1\)-boson funnel region, where \(m_{H_1}\approx 2m_{\tilde N_1}\).  Very large and small values for \(A_0\) and \(\tan\beta\) are disfavoured as they exclude a sneutrino  LSP. These correlations may be somewhat relaxed if non-universal inflaton couplings \(\lambda\) and \(\rho\) are considered.

\section{Sneutrino Relic Density}
\begin{figure}
\begin{minipage}{\columnwidth}
\centering
\includegraphics[clip,width=1\textwidth]{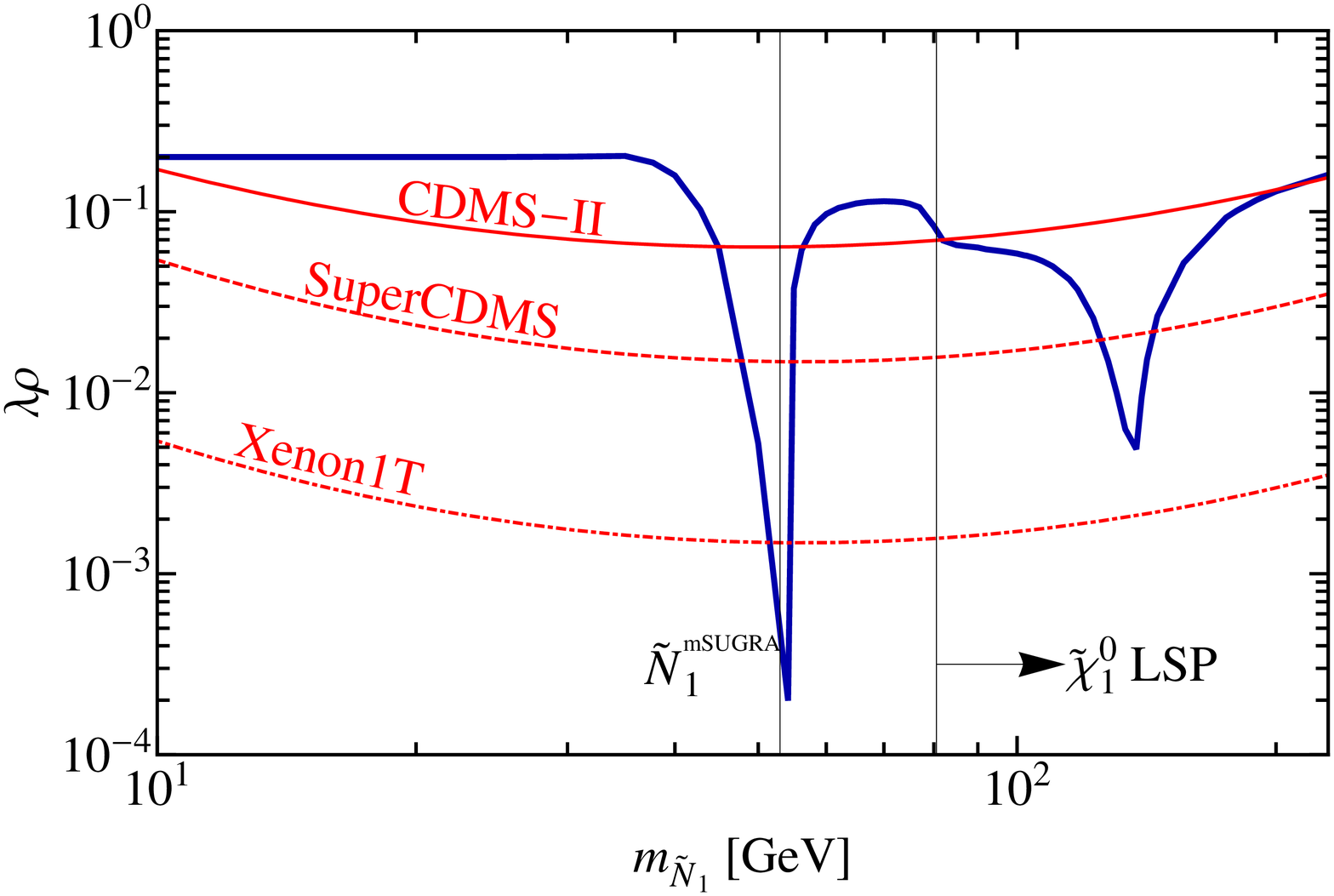}
\end{minipage}
\caption{Effective  annihilation  coupling \(\lambda\rho\) as a function of the LRHS mass \(m_{\tilde  N_1}\) for a relic density \(\Omega_{\rm DM}h^2=0.11\) (blue curve) in the mSUGRA Scenario, Eq.~\ref{ScenII}. The actual sneutrino and neutralino masses in the scenario are indicated by vertical lines. The red curves denote the upper bound on \(\lambda\rho\) as obtained by the  CDMS-II experiment and as expected by the projected sensitivities of SuperCDMS and Xenon1T.}
\label{fig:cdm}
\end{figure}

As was first observed in~\cite{GP}, there exists a new quartic coupling between right-handed sneutrinos and Higgs fields in the $F_D$-term model described by the Lagrangian
\begin{equation}\label{Llsp}
	{\cal L}^{\rm LSP}_{\rm int} =
	\frac{1}{2}\lambda\rho \tilde N^*_i \tilde N^*_i H_uH_d 
	+{\rm H.c.}\; ,
\end{equation}
resulting from the $F$-term of the inflaton field. This leads to several channels of sneutrino annihilation into Higgses, gauge bosons and fermions via the direct quartic coupling as well as $s$-channel Higgs and $t/u$-channel sneutrino exchange once one of the Higgs fields in Eq.~\ref{Llsp} acquires a VEV~\cite{Deppisch:2008bp}. The most effective annihilation process is $b\bar b$ production in the lightest Higgs boson resonance, $m_{\tilde N_1}=m_{H_1}/2$. In order to compute the sneutrino relic density remaining after the sneutrino freezes out of thermal equilibrium and to analyze the constraints on the  effective annihilation coupling \(\lambda\rho\), we adopt the mSUGRA scenario
\begin{eqnarray}
	\label{ScenII}
	&&m_0       = 125 \textrm{ GeV},\ 
	m_{1/2}   = 212 \textrm{ GeV},     \nonumber\\
	&&A_0       = 300 \textrm{ GeV},\ 
	\tan\beta = 30,\ 
	\mu       = 263\textrm{ GeV}\; ,
\end{eqnarray}
while keeping the  LRHS  mass as a free  parameter. The  effective annihilation  coupling \(\lambda\rho\), Eq.~\ref{Llsp}, is then calculated so as to obtain a sneutrino relic density \(\Omega_{\rm DM}\, h^2=0.11\), consistent with observations. Numerical  estimates  of  the allowed  parameters  in  the $(m_{\tilde{N}_1},\lambda\rho)$-plane are shown in Figure~\ref{fig:cdm}. In  order  to account  for  the observed  DM  relic  abundance in  the $H_1$-boson funnel region,  where \(m_{\tilde N_1}\approx m_{H_1}/2\), the effective coupling $\lambda\rho$ should be 
\begin{equation}
	\label{CDMlimits}
	\lambda\rho\ \gtrsim\ 2\times 10^{-4}\; .
\end{equation}
This has to be compared with the upper limit derived from successful inflation ~\cite{Deppisch:2008bp},
\begin{equation}
	\label{Inflimits}
	\lambda\rho\ \lesssim \ 2.3 (5.8)\times 10^{-4}\, ,
\end{equation}
for a minimal (next-to-minimal) K\"ahler potential. In general, we find that LRHS masses larger than about 100~GeV are not possible within a mSUGRA realization of the $F_D$-term model. This is indicated by the value of the neutralino mass in the given mSUGRA scenario as displayed by vertical line in Figure~\ref{fig:cdm}.

Further  constraints on the $(m_{\tilde{N}_1},\lambda\rho)$-plane may be obtained by taking into account the limits from direct searches of experiments which look for scattering between Weakly Interacting Massive Particles and nuclei. Upper limits on \(\lambda\rho\) are  derived by comparing the calculated elastic cross section between the sneutrino LSP and the nucleon with the current bound on the spin-independent nucleon cross section from the CDMS-II experiment and the expected sensitivities of the SuperCDMS extension~\cite{Ogburn:2006hk} and the Xenon1T experiment~\cite{Aprile:2006nz}. These limits are included in Figure~\ref{fig:cdm}. The current bound already excludes large parts    of the $(m_{\tilde{N}_1},\lambda\rho)$-parameter plane. The upgraded experiment SuperCDMS and the proposed Xenon1T experiment will cover most of the parameter space, but will leave open the lightest Higgs-boson  pole  region  which is theoretically favoured by inflation within the mSUGRA framework.

\section{Conclusion}
In this report we discussed the possibility of a right-handed sneutrino LSP as candidate for dark matter in the supersymmetric $F_D$-term model of hybrid inflation. By virtue of a new quartic coupling with the Higgs fields the sneutrino $\widetilde{N}_{\rm LSP}$ can  efficiently annihilate via the lightest Higgs-boson resonance $H_1$ into  pairs of  $b$-quarks, in the kinematic region $m_{H_1} \approx 2 m_{\widetilde{N}_{\rm  LSP}}$, and so drastically reduce its relic density  to the observed value $\Omega_{\rm DM}\,h^2 \approx 0.11$. It might seem that to obtain this particular relation between the masses of the $H_1$ boson and $\widetilde{N}_{\rm LSP}$, a severe tuning of the model parameters is required. However, it is worth stressing here that such a mass relation may easily be achieved within a mSUGRA framework of the $F_D$-term model that successfully realizes hybrid inflation.

Relatively large couplings are required for the sneutrino DM scenario, that could make Higgs bosons decay invisibly, e.g.~$H\to 2\widetilde{N}_{\rm LSP}$.  Also, right-handed sneutrinos could be present in the cascade decays of the heavier supersymmetric particles. The $F_D$-term hybrid inflationary model therefore gives rise to rich phenomenology which  can be probed at high-energy colliders~\cite{NprodLHC}, as  well as in low-energy experiments of lepton flavour and number violation~\cite{CL}.

\end{document}